\documentclass{kluwer}    % Specifies the document style.
\usepackage{epsfig,epsf,rotating}
\newdisplay{guess}{Conjecture}
\newcommand{\beq}{\begin{equation}}
\newcommand{\eeq}{\end{equation}}
\def\be{\begin{equation}}
\def\ee{\end{equation}}
\def\eq#1{{Eq.~(\ref{#1})}}
\newcommand{\as}{\alpha_s}

\newcommand{\bra}[1]{\langle #1 |}
\newcommand{\ket}[1]{|#1\rangle}

\newcommand{\T} {\mbox{T}}

\begin{document}                                                                                   
\begin{article}
\begin{opening}         
\title{\hskip9cm BNL-NT-02/5 \\
\\
Classical Chromo--Dynamics of \\
Relativistic Heavy Ion Collisions\thanks{Lectures given at Cargese Summer School on QCD Perspectives on Hot and Dense Matter, Cargese, France, 6-18 Aug 2001.}} 
\author{Dmitri \surname{Kharzeev}}  
\runningauthor{Dmitri Kharzeev}
\runningtitle{Classical Chromodynamics of Relativistic Heavy Ion Collisions}
\institute{Physics Department, Brookhaven National Laboratory, Upton, NY11973-5000}
\date{March 15, 2002}

\begin{abstract}
Relativistic heavy ion collisions produce thousands of particles, and it is sometimes difficult to believe 
that these processes allow for a theoretical description directly in terms of the underlying theory 
-- QCD. 
However once the parton densities are sufficiently large, an essential simplification occurs -- the 
dynamics becomes semi--classical. As a result, a simple {\it ab initio} approach to the 
nucleus--nucleus collision dynamics may be justified. In these lectures, we describe the application of 
these ideas to the description of multi--particle production in relativistic heavy ion collisions. 
We also discuss the r{\^o}le of semi--classical fields in the QCD vacuum in hadron 
interactions at low and high energies.  

\end{abstract}

\end{opening}           

\section{What is Chromo--Dynamics?}  
                    % Produces section heading.  Lower-level
                    % sections are begun with similar 
Strong interaction is, indeed, the strongest force of Nature. 
It is responsible for over $80 \%$ of the baryon masses, and thus 
for most of the mass of everything on Earth 
and in the Universe. Strong interactions bind  
nucleons in nuclei, which, being then bound into molecules by 
much weaker electro-magnetic forces, 
give rise to the variety of the physical World.
Quantum Chromo--Dynamics is {\it the} theory of strong interactions, 
and its practical importance is thus undeniable. 
But QCD is more than a useful tool -- it is a consistent and 
very rich field theory, which continues to serve as a stimulus for, 
and testing ground of, many exciting ideas and new methods in 
theoretical physics. 

These lectures will deal with QCD of strong color fields, which can be explored 
in relativistic heavy ion collisions. (See the lectures by E. Iancu, A. Leonidov, L. McLerran \cite{ILM}, 
A.H. Mueller \cite{AHM}, and R. Venugopalan in this volume for complementary presentation of the subject and 
more details.)

\subsection{QCD: the Lagrangean}

So what is QCD? From the early days of the accelerator experiments 
it has become clear that the number of hadronic resonances 
is very large, suggesting that all hadrons may be 
classified in terms of a smaller number of (more) 
fundamental constituents. A convenient classification was offered 
by the quark model, but QCD was not born until the hypothetical 
existence of quarks was not supplemented by the principle of 
local gauge invariance, previously established as the basis 
of electromagnetism.
The resulting Lagrangian has the form 
\beq
{\cal{L}} = -{1 \over 4} G_{\mu\nu}^a G_{\mu\nu}^a + \sum_f \bar{q}_f^a 
(i \gamma_{\mu} D_{\mu} - m_f) q_f^a; \label{lagr}
\eeq
the sum is over different colors $a$ and quark flavors $f$; 
the covariant derivative is $D_{\mu} = \partial_{\mu} - i g 
A_{\mu}^a t^a$, where $t^a$ is the generator of the color group $SU(3)$, $A_{\mu}^a$ 
is the gauge (gluon) field and $g$ is the coupling constant. 
The gluon field strength tensor is given by 
\beq
G_{\mu\nu}^a = \partial_{\mu} A_{\nu}^a - \partial_{\nu} A_{\mu}^a + 
g f^{abc} A_{\mu}^b A_{\nu}^c, \label{ftens}
\eeq
where $f^{abc}$ is the structure constant of $SU(3)$: 
$[t^a, t^b] = i f^{abc} t^c$. 

\subsection{Asymptotic freedom}

Due to the quantum effects of vacuum polarization, the charge  
in field theory can vary with the distance. In electrodynamics, summation 
of the electron--positron loops in the photon propagator leads 
to the following expression for the effective charge, valid 
at $r \gg r_0$: 
\beq
\alpha_{em}(r) \simeq {3 \pi \over 2 \ln(r/r_0)}. \label{aem}
\eeq
This formula clearly exhibits the ``zero charge'' problem \cite{Landau} 
of QED: in the local 
limit $r_0 \to 0$ the effective charge vanishes at any finite 
distance away from the bare charge due to the screening. 
Fortunately, because of the smallness of the physical coupling, 
this apparent inconsistency of the theory 
manifests itself only at very short distances 
$\sim exp\{-3\pi/[2 \alpha_{em}]\}, \ \alpha_{em} \simeq 1/137$. 
Such short distances are (and probably will 
always remain) beyond the reach of experiments, and one can 
safely use QED as a truly effective theory.

As it has been established long time ago \cite{asfr}, QCD is drastically 
different from electrodynamics in possessing the remarkable property 
of ``asymptotic freedom'' -- due to the 
fact that gluons carry color, the behavior 
of the effective charge $\alpha_s = g^2/4 \pi$ 
changes from the familiar from QED screening to anti--screening:
\beq 
\alpha_{s}(r) \simeq {3 \pi \over (11 N_c / 2 - N_f) \ln(r_0/r)}; 
\label{as} 
\eeq 
as long as the number of flavors does not exceed $16$ ($N_c = 3$), 
the anti--screening 
originating from gluon loops overcomes the screening due to quark--antiquark 
pairs, and the theory, unlike electrodynamics, is weakly coupled 
at short distances: $ \alpha_{s}(r) \to 0$ when $r \to 0$. 

\subsection{Chiral symmetry}

In the limit of massless quarks, QCD Lagrangian (\ref{lagr}) possesses 
an additional symmetry $U_L(N_f) \times U_R(N_f)$ 
with respect to the independent transformation of left-- and right--handed 
quark fields $q_{L,R} = {1 \over 2}(1 \pm \gamma_5) q$: 
\beq
q_L \to V_L q_L; \ \ q_R \to V_R q_R; \ \ V_L, V_R \in U(N_f); \label{chiral}
\eeq
this means that left-- and right--handed quarks are not correlated.  
Even a brief look into the Particle Data tables, or simply in the 
mirror, can convince anyone 
that there is no symmetry between left and right in the physical World. 
One thus has to assume that the symmetry (\ref{chiral}) is spontaneously 
broken in the vacuum.  The flavor composition of the existing eight Goldstone 
bosons (3 pions, 4 kaons, and the $\eta$) suggests that the 
$U_A(1)$ part of $U_L(3) \times U_R(3) = 
SU_L(3) \times SU_R(3) \times U_V(1) \times U_A(1)$ does not exist.  
This constitutes the famous ``$U_A(1)$ problem''. 

\subsection{The origin of mass}

There is yet another problem with the chiral limit in QCD. Indeed, as the 
quark masses are put to zero, the Lagrangian (\ref{lagr}) does not contain 
a single dimensionful scale -- the only parameters are pure numbers $N_c$ 
and $N_f$. The theory is thus apparently 
invariant with respect to scale transformations, 
and the corresponding scale current is conserved: 
$\partial_{\mu} s_{\mu} = 0$.      
However, the absence of a mass scale would imply that all physical 
states in the theory should be massless!

\subsection{Quantum anomalies and classical solutions}

Both apparent problems -- the missing $U_A(1)$ symmetry and the 
origin of hadron masses -- are related to quantum anomalies. 
Once the coupling to gluons is included, both flavor singlet axial 
current and the scale current
cease to be conserved; their divergences 
become proportional to the $\alpha_s G_{\mu\nu}^a \tilde{G}_{\mu\nu}^a$ and 
$\alpha_s G_{\mu\nu}^a G_{\mu\nu}^a$ gluon operators, correspondingly. 
This fact by itself would not have dramatic consequences 
if the gluonic vacuum were ``empty'', with $G_{\mu\nu}^a = 0$. 
However, it appears that due to non--trivial topology of the $SU(3)$ 
gauge group, QCD equations of motion allow classical solutions even 
in the absence of external color source, i.e. in the vacuum. 
The well--known example of a classical solution is the instanton, 
corresponding 
to the mapping of a three--dimensional sphere $S^3$ into the $SU(2)$ subgroup 
of $SU(3)$; its existence was shown to solve the $U_A(1)$ problem.

\subsection{Confinement}

The list of the problems facing us in the study of QCD would not be complete 
without the most important problem of all -- why are the colored quarks and 
gluons excluded from the physical spectrum of the theory? 
Since confinement does not appear in perturbative treatment of the theory, 
the solution of this problem, again, must lie in the properties of the 
QCD vacuum.

\subsection{Understanding the Vacuum}

As was repeatedly stated above, the most important problem facing us in 
the study of all aspects of QCD is understanding the structure of the 
vacuum, which, in a manner of saying, does not at all behave as 
an empty space, but as a physical entity with a complicated structure. 
As such, the vacuum can be excited, altered and modified in physical 
processes \cite{td}.
   
\section{Strong interactions at short and large distances}

In this lecture we will investigate the influence of QCD vacuum on hadron 
interactions at short and large distances. To make the problem treatable, 
we will limit ourselves to heavy quarkonia. 
In this lecture I will describe two 
recent results -- one on the scattering of heavy quarkonia at very low energies, 
another on high--energy scattering. The common idea behind these two examples is 
to explore the influence of the QCD vacuum on hadron interactions.   
The presentation will be schematic, and 
I refer the interested reader to the original papers \cite{FK} and \cite{KL} for  
details. 

\subsection{The long--range forces of QCD}
\vskip0.3cm
\subsubsection{Perturbation theory}

Let us begin with a somewhat academic problem -- the scattering of two heavy 
quarkonium states at very low energies. 
The Wilson operator product expansion allows one to write down 
the scattering amplitude (in the Born approximation) of two small color dipoles
in the following form\cite{Peskin}:
\begin{eqnarray}
V(R) &=& -i \int dt  
\bra{0} \T \left( \sum_i c_i O_i (0)\right) \left( \sum_j c_j O_j (x)\right) 
\ket{0}, 
\label{ope1}
\end{eqnarray}
where $x = (t, R)$, $O_i(x)$ is the set of local gauge-invariant operators expressible 
in terms of gluon fields, and $c_i$ are the coefficients which reflect 
the structure of the color dipole.
At small (compared to the binding energy of the dipole) energies, 
the leading operator in (\ref{ope1}) is the square of the chromo-electric 
field $(1/2) g^2 {\bf E}^2$ \cite{Voloshin,Peskin}. 
Keeping only this leading operator, we can rewrite (\ref{ope1}) in
a simple form   
\begin{eqnarray}
V(R) &=& 
- i\Big ( \bar d_2 \frac{a_0^2}{\epsilon_0} \Big)^2 
\int d t  
\bra{0} \T \ \frac{1}{2}g^2 {\bf E}^2 (0)\ 
\frac{1}{2} g^2 {\bf E}^2(t,R) \ket{0}, 
\label{pot1}
\end{eqnarray}
where $\bar d_2$ is the corresponding Wilson coefficient defined by
\begin{equation}
\bar d_2 \frac{a_0^2}{\epsilon_0}
=\frac{1}{3N}\bra{\phi} r^i \frac{1}{H_a + \epsilon} r^i \ket{\phi},
\end{equation}
where we 
have explicitly factored out the dependence on the quarkonium Bohr 
radius $a_0$ and the Rydberg energy $\epsilon_0$; $N$ is the number 
of colors, and $\ket{\phi}$ is the quarkonium wave function, which 
is Coulomb in the heavy quark limit\footnote{The Wilson 
coefficients $\bar d_2$, 
evaluated in the large $N$ limit, are available for $S$ \cite{Peskin} and 
$P$ \cite{DK} quarkonium states.}. 
In physical terms, the structure of (\ref{pot1}) is transparent: 
it describes elastic scattering of two dipoles which act on 
each other by chromo-electric dipole fields; color neutrality permits 
only the square of dipole interaction. 
It is convenient to express $g^2 {\bf E}^2$ in terms of the 
gluon field strength tensor \cite{NS}:
$$
g^2 {\bf E}^2 =
- \frac{1}{4}g^2    G_{\alpha\beta}G^{\alpha\beta}
+g^2(- G_{0\alpha} G_0^\alpha
+\frac{1}{4} g_{00} G_{\alpha\beta}G^{\alpha\beta})
=
$$
\begin{eqnarray}
= \frac{8 \pi^2}{b} \theta_\mu^\mu + g^2 \theta_{00}^{(G)}  
 \label{e2}
\end{eqnarray}
where
\begin{eqnarray}
\theta_\mu^\mu \equiv
\frac{\beta(g)}{2g} G^{\alpha\beta a} G_{\alpha\beta}^{a} =
- \frac{b g^2}{32 \pi^2} G^{\alpha\beta a} G_{\alpha\beta}^{a} 
 \ . \label{trace}
\end{eqnarray}
Note that as a consequence of scale anomaly, 
$\theta_\mu^\mu$ is the trace of the energy-momentum tensor 
of QCD in the chiral limit of vanishing light quark masses. 

Let us now introduce the spectral representation for the correlator 
of the trace of energy-momentum tensor:
\begin{eqnarray}
\bra{0} \T \theta_\mu^\mu(0) \theta_\mu^\mu(x) \ket{0} = 
\int d \sigma^2 \rho_\theta (\sigma^2) \Delta_F(x;\sigma^2),
\label{spectral}
\end{eqnarray}
where $\rho_\theta (\sigma^2)$ is the spectral density 
and $\Delta_F(x;\sigma^2)$ is the Feynman propagator of a scalar field.
Using the representation (\ref{spectral}) in (\ref{pot1}), we get
\begin{eqnarray}
V_\theta(R) &=& 
 -
\Big ( \bar d_2 \frac {a_0^2}{\epsilon_0} \Big )^2 
\Big ( \frac{4 \pi^2 }{b}\Big ) ^2
\int d \sigma^2  \rho_\theta (\sigma^2)
\frac{1}{4\pi R}e^{-\sigma R}.
\label{yukawa}
\end{eqnarray}
The potential (\ref{yukawa}) is simply a 
superposition of Yukawa potentials
corresponding to the exchange of scalar quanta of mass $\sigma$.

Our analysis so far has been completely general; the dynamics 
enters through the spectral density. In perturbation theory, 
for $SU(N)$, one has
\begin{eqnarray}
\rho^{\rm pt}_{\theta}(q^2) &=& 
\left ( \frac{bg^2}{32\pi^2} \right )^2\frac{N^2-1}{4\pi^2} q^4  .
\label{pdens}
\end{eqnarray}
Substituting (\ref{pdens}) into (\ref{yukawa}) and performing 
the integration over invariant mass $\sigma^2$, we get, for 
$N=3$
\begin{eqnarray}
V_\theta(R) &=&
-  g^4 \Big ( \bar d_2 \frac{a_0^2}{\epsilon_0}\Big)^2
\frac{15}{8\pi^3}\frac{1}{R^7}.
\label{casimir}
\end{eqnarray} 

The $\propto R^{-7}$ dependence of the potential (\ref{casimir}) is a 
classical result known from atomic physics \cite{cas}; as is apparent in 
our derivation (note the time integration in (\ref{yukawa})), 
the extra $R^{-1}$ as compared to the 
Van der Waals potential $\propto R^{-6}$ is the consequence of the fact that 
the dipoles we consider fluctuate in time, and the characteristic 
fluctuation time $\tau \sim \epsilon_0^{-1}$, 
is small compared to the spatial separation of the ``onia'' : $\tau \ll R$.

Let us note finally that the second term in (\ref{e2}) gives the contribution of the same
order in $g$; this contribution is due to the tensor $2^{++}$ state of 
two gluons and can be evaluated in a completely analogous way.
Adding this contribution to (\ref{casimir}), changes the factor of $15$ in (\ref{casimir}) to $23$, and  
we reproduce the result of ref.~\cite{Peskin}, which shows the equivalence of the 
spectral representation method used here and the functional method of ref. \cite{Peskin}.

\begin{figure}[tb]
\epsfxsize=0.5\textwidth
\centerline{\epsffile{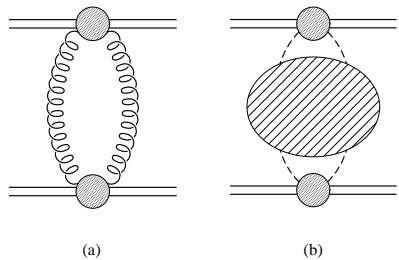}}
\caption{Contributions to the scattering amplitude from (a) two
gluon exchange and (b) correlated two pion exchange.}     
\end{figure}

\subsubsection{Beyond the perturbation theory: scale anomaly and the role of pions}

At large distances, the perturbative description breaks down, because, 
as can be clearly seen from (\ref{yukawa}), 
the potential becomes determined by the spectral density at small $q^2$, 
where the transverse momenta of the gluons become small. 
At small invariant masses, we have therefore to saturate the physical 
spectral density by the lightest state
allowed in the scalar channel -- two pions. 
Since, according to (\ref{trace}), $\theta^\alpha_\alpha$ is 
gluonic operator, this requires the knowledge of the coupling of gluons to pions. 
This looks like a hopeless non--perturbative problem, but it can nevertheless 
be rigorously solved, as it was shown in ref. \cite{VZ} 
(see also \cite{NS}). The idea is the following:  
at small 
pion momenta, the energy--momentum tensor can be accurately computed 
using the low--energy chiral Lagrangian:
\begin{equation}
\theta_\mu^\mu =
-\partial_\mu \pi^a \partial^\mu\pi^a +2 m_\pi^2 \pi^a \pi^a + \cdots 
\label{trl}
\end{equation}
Using this expression, in the chiral limit 
of vanishing pion mass one gets an elegant result \cite{VZ} 
\begin{equation}
\bra{0} \frac{\beta(g)}{2g} G^{\alpha\beta a} G_{\alpha\beta}^{a} 
\ket {\pi^+\pi^-} = q^2. \label{vz}
\end{equation}
Now that we know the coupling of gluons to the two pion state, 
the pion--pair contribution to the spectral density can be 
easily computed by performing the simple phase space integration, 
with the result  
\begin{equation}
\rho^{\pi\pi}_\theta(q^2) = \frac{3}{32\pi^2} q^4 , \label{ppi}
\end{equation}  
which leads to the following long--distance potential \cite{FK}:
\begin{equation}
V^{\pi \pi}(R) 
\rightarrow 
-\Big (\bar d_2 \frac{a_0^2}{\epsilon_0}\Big )^2
 \left ( \frac{4 \pi^2}{b} \right )^2
\frac{3}{2} (2 m_\pi)^4 
\frac{m_\pi^{1/2}}{(4\pi R) ^ {5/2} } 
e^{-2m_\pi R}
\qquad \mbox{as }R\rightarrow \infty . \label{uncor}
\end{equation}
Note that, unlike the perturbative result  
which is manifestly $\sim g^4$, the amplitude (\ref{uncor}) 
is $\sim g^0$ -- this ``anomalously" strong interaction 
is the consequence of scale anomaly\footnote{Of course, in the 
heavy quark limit the amplitude (\ref{uncor}) will nevertheless vanish, 
since $a_0 \to 0$ and $\epsilon_0 \to \infty$.}.   

While the shape of the potential in general depends on the spectral density, 
which is fixed theoretically only at relatively small invariant mass, the overall 
strength of the non-perturbative interactions is fixed by low energy theorems 
and is determined by the energy density of QCD vacuum.  
Indeed, in the heavy quark limit, one can derive the following sum rule \cite{FK}:
\be    
\int d^3{\bf R} \left( V(R) - V^{pt}(R) \right) = 
\Big ( \bar d_2 \frac {a_0^2}{\epsilon_0} \Big )^2 
\Big ( \frac{4 \pi^2 }{b}\Big ) ^2 \ 16 \left|\epsilon_{vac}\right|, 
\label{sr2}
\ee
which expresses the overall strength of the interaction between 
two color dipoles in terms of the energy density of the 
non-perturbative QCD vacuum. 

\subsubsection{Does $\alpha_s$ ever get large?}   

Asymptotic freedom ensures the applicability of QCD perturbation theory 
to the description of processes accompanied by high momentum transfer $Q$. 
However, as $Q$ decreases, the strong coupling $\alpha_s(Q)$ 
grows, and the convergence of perturbative series is lost. 
How large can $\alpha_s$ get? 
The analyzes of many observables suggest that the QCD coupling 
may be ``frozen'' in the infrared region at the value 
$\left<\alpha_s\right>_{IR} \simeq 0.5$ (see 
\cite{Yuri} and references therein). 
Gribov's program \cite{Gribov} relates the freezing of the coupling constant 
to the existence of massless quarks, which leads to the ``decay'' of the vacuum at 
large distances similar to the way it happens in QED in the presence of 
``supercritical'' charge $Z > 1/\alpha$. One may try to  
infer the information about 
the behavior of the coupling constant at large distances by 
performing the matching of the fundamental theory onto the effective chiral 
Lagrangian at a scale $Q \simeq 4 \pi f_{\pi} \simeq 1$ GeV, at which the ranges of validity 
of perturbative and chiral descriptions meet \cite{FK}. 
It is easy to see that in the chiral limit the matching of the potentials (\ref{uncor}) 
and (\ref{casimir}) yields\footnote{The matching procedure of course can be performed directly 
for the correlation function of the energy--momentum tensor.} 
the coupling constant which freezes at the  
value 
\beq
\langle \alpha_s \rangle_{IR} = 
{6 \sqrt{2}\ \pi \over 11 N_c - 2 N_f} \ \sqrt{{N_f^2-1 \over N_c^2-1}}; \label{frozal}
\eeq 
numerically, for QCD with $N_c=3$ and $N_f=2$ one finds 
$\langle \alpha_s \rangle_{IR} \simeq 0.56$.  
Note that this expression has an expected $N_c$ dependence in the topological expansion limit of 
$N_c \to \infty, N_f/N_c = const$.

Since the trace of the energy momentum tensor in general relativity is linked to the curvature 
of space--time, the matching procedure leading to Eq. (\ref{frozal}) has an interesting geometrical 
interpretation: it corresponds to the matching, at a relatively large distance, of curved space--time 
of the fundamental QCD with the flat space--time of the chiral theory.  

\subsection{High--energy scattering: scale anomaly and the ``soft'' Pomeron}

In a 1972 article entitled ``Zero pion mass limit in interaction 
at very high energies" \cite{AG}, A.A. Anselm and V.N. Gribov 
posed an interesting question: what is the total cross section of 
hadron scattering in the chiral limit of $m_{\pi} \to 0$?  
On one hand, as everyone believes since 
the pioneering work of H. Yukawa, the range of strong interactions is 
determined by the mass of the lightest meson, i.e. is proportional to  
$\sim m_{\pi}^{-1}$. The total cross sections may then be expected 
to scale as $\sim m_{\pi}^{-2}$, and would tend to infinity as 
$m_{\pi} \to 0$. On the other hand, soft--pion 
theorems, which proved to be 
very useful in understanding low--energy hadronic phenomena, 
state that hadronic amplitudes do not possess singularities 
in the limit $m_{\pi} \to 0$, and 
one expects that the theory must remain   
self-consistent in the limit of the vanishing pion mass.        
At first glance, the advent of QCD has not made this 
problem any easier; on the contrary, the presence of massless gluons in 
the theory apparently introduces another long--range interaction. 
Here, we will try to address this problem 
considering the scattering of small color dipoles. 

Again, perturbation theory provides a natural starting point.
In the framework of perturbative QCD, 
a systematic approach to high energy scattering was developed by Balitsky, Fadin, Kuraev 
and Lipatov \cite{BFKL}, who demonstrated that the ``leading 
log'' terms in the scattering amplitude 
of type $(g^2 ln\ s)^n$ (where $g$ is the strong 
coupling) can be re-summed, giving rise to the so--called ``hard'' 
Pomeron. Diagramatically, BFKL equation 
describes the $t-$channel exchange of ``gluonic ladder''  
-- a concept familiar 
from the old--fashioned multi--peripheral model. 

It has been found, however, that at sufficiently high 
energies the perturbative description breaks down \cite{Mueller1}, \cite{Dok}. 
The physical reason for this is easy to understand: the higher 
the energy, the larger impact parameters contribute to the scattering, and 
at large transverse distances the perturbation theory inevitably fails, since  
the virtualities of partons in the ladder diffuse to small values. 
%(These arguments were formulated in QCD in terms of the 
%operator product expansion, which was found to break down at high 
%energies \cite{Mueller1}). 
%The origin of this difficulty can be traced back to the break--down 
%of scale invariance \cite{scale}in QCD, induced by the renormalization. 
 At this point, the 
following questions arise: 
Does this mean that the problem 
becomes untreatable? 
Does the same difficulty appear at large distances in low--energy scattering?  
And, finally, what (if any) is the role played by pions?

The starting point of the approach proposed in \cite{KL}   
is the following: 
among the higher order, $O(\alpha^2_S)$ ($\alpha_S= g^2/4\pi$) ,
corrections to the BFKL kernel one can 
isolate a particular class of diagrams which include the propagation 
of two gluons in the scalar color singlet channel $J^{PC}=0^{++}$. 
We then show that, as a consequence 
of scale anomaly, these, apparently  $O(\alpha^2_S)$, contributions become 
the {\it dominant} ones, $O(\alpha^0_S)$. This is similar to our previous discussion 
in Sect. 2.1, where the interaction potential, proportional to $\alpha_S^2$, at large 
distances turned into a ``chiral'' potential $\sim \alpha_S^0$ due to the scale anomaly.   

One way of understanding the disappearance of the coupling constant 
in the spectral density of the $g^2 G^2$ operator is to assume that the 
non-perturbative QCD vacuum is dominated by the semi--classical fluctuations 
of the  gluon field. Since the strength of the classical gluon field 
is inversely proportional to the coupling, $G \sim 1/g$, 
the quark zero modes, and the spectral density of their pionic excitations, 
appear independent of the coupling constant.

The explicit calculation using the methods of \cite{GLR} yields 
the   
power--like
behavior of the total cross section:
\beq \label{10}
\sigma_{tot} = \sum^{\infty}_{n = 0}\, \sigma_n \, =\,\sigma^{BORN} s^{\Delta}
\,\,,
\eeq
where $\sigma^{BORN}$ is the cross section due to two gluon exchange, and
the non--perturbative contribution to the intercept $\Delta$ is \cite{KL}
\beq \label{14}
\Delta \,\,=\,\frac{\pi^2}{2} \times \left( \frac{8 \pi}{b}
\right)^2\,\times \frac{18}{32 \pi^2}\,\int\,\frac{ d M^2}{M^6}
\,\left(\,\rho^{phys}_{\theta}( M^2 )\,\,-\,\,\rho^{pQCD}_{\theta}( M^2
)\,\right)\,\,.
\eeq
Using the chiral formula (\ref{ppi}) for $\rho^{phys}_{\theta}$ for $M^2 < M_0^2$, 
 we obtain the following result \cite{KL}:
\beq \label{16}
\Delta\,\,=\,\, \frac{1}{48}\,\,\ln \,\frac{M^2_0}{4 m^2_{\pi}}\,\,.
\eeq
The precise value of the 
matching scale $M^2_0$ as extracted from the low--energy theorems                       
depends somewhat on detailed form of the spectral density, 
and can vary within the range of $M^2_0 = 4 \div 
6\,$GeV$^2$. 
Fortunately, the dependence of \eq{16} on $M_0$ is only 
logarithmic, and varying it in this range leads to 
\beq
\Delta = 0.08 \div 0.1, \label{number} 
\eeq
in agreement with the phenomenological intercept of the ``soft'' 
Pomeron, $\Delta \simeq 0.08$. 

At present, the language used in the description of hadron interactions at low and  
high energies is very different. Yet, as the two examples discussed above imply, 
both limits may appear to be determined by the same fundamental object -- 
the QCD vacuum.

\section{QCD in the classical regime }

Most of the applications of QCD so far have been  
limited to the short distance regime of high momentum transfer, 
where the theory becomes weakly coupled and can be linearized.
While this is the only domain where our theoretical tools based 
on perturbation theory are adequate, this is also the domain in 
which the beautiful non--linear structure of QCD does not yet reveal 
itself fully. On the other hand, as soon as we decrease the momentum 
transfer in a process, the dynamics rapidly becomes non--linear, but our  
understanding is hindered by the large coupling. 
Being perplexed by this problem, one is 
tempted to dream about an environment in which the coupling is weak, 
allowing a systematic theoretical treatment, but the fields are strong, 
revealing the full non--linear nature of QCD. 
I am going to argue now that this environment can be created on Earth 
with the help of relativistic heavy ion colliders.    
Relativistic heavy ion collisions allow to probe QCD in the non--linear 
regime of high parton density and high color field strength.

\begin{figure}[h]
\begin{minipage}[t]{140mm}
\includegraphics[width=21pc]{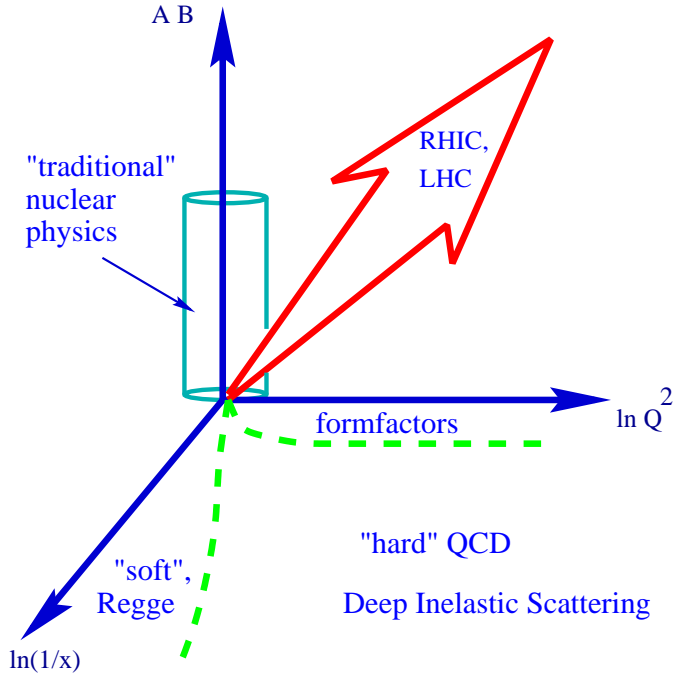}
\caption{The place of relativistic heavy ion physics in the study 
of QCD; the vertical axis is the product of atomic numbers of projectile 
and target, and the horizontal axes are the momentum 
transfer $Q^2$ and rapidity $y = \ln(1/x)$ ($x$ is the Bjorken scaling 
variable).}
\label{fig:rhic}
\end{minipage}
\end{figure}

It has been conjectured long time ago that the dynamics of QCD  
in the high density domain  may become qualitatively different: in parton language, 
this is best described in terms of {\it parton saturation} \cite{GLR,MQ,BM}, and in the language 
of color fields -- in terms of the {\it classical} Chromo--Dynamics \cite{MV}; see the lectures 
\cite{ILM} and \cite{AHM} and references therein. 
In this high density regime,  
the transition amplitudes are dominated not by quantum fluctuations, but by 
the configurations of classical field containing large, $\sim 1/\alpha_s$, 
numbers of gluons. One thus uncovers new 
non--linear features of QCD, 
which cannot be investigated in the more traditional applications
based on the perturbative approach. 
The classical color fields in the initial nuclei (the 
``color glass condensate'' \cite{ILM}) can be thought of as 
either perturbatively generated, or 
as being a topologically non--trivial superposition of the Weizs{\"a}cker-Williams 
radiation and the quasi--classical vacuum fields \cite{inst,inst1,KKL}.    

\subsection{Geometrical arguments}
 
Let us consider an external probe $J$ interacting with the 
nuclear target of atomic number $A$. At small values of Bjorken $x$, 
by uncertainty principle the interaction develops over large 
longitudinal distances $z \sim 1/mx$, where $m$ is the 
nucleon mass. As soon as $z$ becomes larger than the nuclear diameter, 
the probe cannot distinguish between the nucleons located on the front and back edges 
of the nucleus, and all partons within the transverse area $\sim 1/Q^2$ 
determined 
by the momentum transfer $Q$ participate in the interaction coherently. 
The density of partons in the transverse plane is given by
\beq
\rho_A \simeq {x G_A(x,Q^2) \over \pi R_A^2} \sim A^{1/3},
\eeq
where we have assumed that the nuclear gluon distribution  
scales with the number of nucleons $A$. The probe interacts with 
partons with cross section $\sigma \sim \alpha_s / Q^2$; therefore, 
depending on the magnitude of momentum transfer $Q$, atomic number $A$, 
and the value of Bjorken $x$, one may encounter two regimes:
\begin{itemize}
\item{$\sigma \rho_A \ll 1$ -- this is a familiar ``dilute'' regime of 
incoherent interactions, which is well described by the methods of 
perturbative QCD;}
\item{$\sigma \rho_A \gg 1$ -- in this regime, we deal with a dense 
parton system. Not only do the ``leading twist'' expressions become 
inadequate, but also the expansion in higher twists, i.e. in 
multi--parton correlations, breaks down here.}
\end{itemize} 

\begin{figure}[h]
\begin{minipage}[t]{140mm}
\includegraphics[width=14pc]{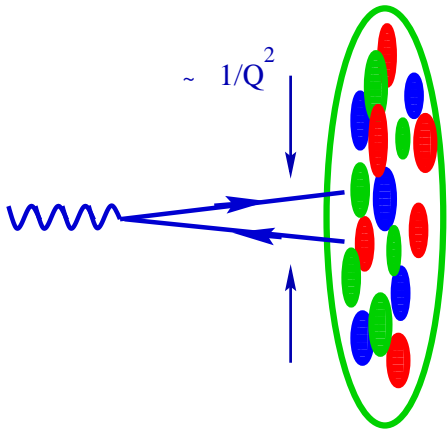}
\caption{Hard probe interacting with the nuclear target 
resolves the transverse distance $\sim 1/\sqrt{Q}$ ($Q^2$ is the square of the 
momentum transfer) and, in the target rest frame, the longitudinal 
distance $\sim 1/(m x)$ ($m$ is the nucleon mass and $x$ -- Bjorken variable).} 

\label{fig:satpict}
\end{minipage}
\end{figure}

The border between the two regimes can be found from the condition 
$\sigma \rho_A \simeq 1$; it determines the critical value of the 
momentum transfer (``saturation scale''\cite{GLR}) at which the parton system 
becomes to look dense to the probe\footnote{Note that since 
$x G_A(x,Q_s^2) \sim A^{1/3}$, 
which is the length of the target, this expression 
in the target rest frame can also be understood as describing a broadening 
of the transverse momentum resulting from the multiple re-scattering 
of the probe.}:
\beq 
Q_s^2 \sim  \alpha_s \ {x G_A(x,Q_s^2) \over \pi R_A^2}. \label{qsat}
\eeq
In this regime, the number of gluons from (\ref{qsat}) is given by 
\beq
x G_A(x,Q_s^2) \sim {\pi \over \alpha_s(Q_s^2)}\ Q_s^2 R_A^2, \label{glustr}
\eeq
where $Q_s^2 R_A^2 \sim A$. 
One can see that the number of gluons 
is proportional to the {\em inverse} of $\alpha_s(Q_s^2)$, and 
becomes large in the weak coupling regime. In this regime,
as we shall now discuss, the dynamics is likely to become 
essentially classical. 

\subsection{Saturation as the classical limit of QCD}

Indeed, the condition (\ref{qsat}) can be derived in the following, 
rather general, way. As a first step, let us re-scale the gluon fields 
in the Lagrangian (\ref{lagr}) as follows: $A_{\mu}^a \to \tilde{A}_{\mu}^a = 
g A_{\mu}^a$. In terms of new fields, $\tilde{G}_{\mu \nu}^a = 
g G_{\mu \nu}^a = \partial_{\mu} \tilde{A}_{\nu}^a - \partial_{\nu} 
\tilde{A}_{\mu}^a +  f^{abc} \tilde{A}_{\mu}^b \tilde{A}_{\nu}^c$, 
and the dependence of the action corresponding to the 
Lagrangian (\ref{lagr}) on the coupling constant is given by  
\beq
S \sim \int {1 \over g^2}\ \tilde{G}_{\mu \nu}^a  \tilde{G}_{\mu \nu}^a 
\ d^4 x. \label{act}
\eeq
Let us now consider a classical configuration of gluon fields; by definition, 
$\tilde{G}_{\mu \nu}^a$ in such a configuration does not depend on 
the coupling, and the action is large, $S \gg \hbar$. The number of 
quanta in such a configuration is then
\beq
N_g \sim {S \over \hbar} \sim {1 \over \hbar \ g^2}\ \rho_4 V_4, \label{numb}
\eeq
where we re-wrote (\ref{act}) as a product of four--dimensional 
action density $\rho_4$ and the four--dimensional volume $V_4$. 
 
Note that since (\ref{numb}) depends only on the product of the Planck constant $\hbar$ and 
the coupling $g^2$, the classical limit $\hbar \to 0$ is indistinguishable from the 
weak coupling limit $g^2 \to 0$. The weak coupling limit of small $g^2 = 4 \pi \alpha_s$ 
therefore corresponds to the semi--classical regime.

The effects of non--linear interactions among the gluons become 
important when $\partial_{\mu} \tilde{A}_{\mu} \sim \tilde{A}_{\mu}^2$ 
(this condition can be made explicitly gauge invariant if we derive it 
from the expansion of a correlation function of gauge-invariant 
gluon operators, e.g., $\tilde{G}^2$). In momentum space, this 
equality corresponds to 
\beq
Q_s^2 \sim \tilde{A}^2 \sim (\tilde{G}^2)^{1/2} = 
\sqrt{\rho_4}; \label{nonlin}
\eeq
$Q_s$ is the typical value of the gluon momentum below which 
the interactions become essentially non--linear. 

Consider now a nucleus $A$ boosted to a high momentum. By uncertainty 
principle, the gluons with transverse momentum $Q_s$ are extended 
in the longitudinal and proper time directions by $\sim 1/Q_s$; 
since the transverse area is $\pi R_A^2$, the four--volume 
is $V_4 \sim \pi R_A^2 / Q_s^2$. The resulting four--density from 
(\ref{numb}) is then 
\beq
\rho_4 \sim \alpha_s\ {N_g \over V_4} \sim \alpha_s\ {N_g\ Q_s^2 
\over \pi R_A^2} 
\sim Q_s^4, \label{class}
\eeq
where at the last stage we have used the non--linearity condition (\ref{nonlin}),  
$\rho_4 \sim Q_s^4$. It is easy to see that (\ref{class}) coincides with the 
saturation condition (\ref{qsat}), since the number of gluons in the 
infinite momentum frame $N_g \sim x G(x,Q_s^2)$. 
\vskip0.3cm

In view of the significance of saturation criterion for the rest of the material 
in these lectures, let us present yet another argument, traditionally followed 
in the discussion of classical limit in electrodynamics \cite{LL}.
The energy of the gluon field per unit volume is $\sim \vec{E}^{a 2}$. The number 
of elementary ``oscillators of the field'', also per unit volume, is $\sim \omega^3$.
To get the number of the quanta in the field we have to divide the energy of the field 
by the product of the number of the oscillators $\sim \omega^3$ and the average energy 
$\hbar \omega$ of the gluon:
\be
N_{{\vec k}} \sim {\vec{E}^{a 2} \over \hbar \omega^4}. \label{qc3}
\ee

The classical approximation holds when $N_{{\vec k}} \gg 1$. Since the energy $\omega$ of the 
oscillators is related to the time $\Delta t$ over which the average energy is computed 
by $\omega \sim 1/\Delta t$, we get 
\be
\vec{E}^{a 2} \gg {\hbar \over (\Delta t)^4}. \label{llcon}  
\ee
Note that the quantum mechanical uncertainty principle for the energy of the field 
reads 
\be
\vec{E}^{a 2} \  \omega^4 \sim \hbar,
\ee
so the condition (\ref{llcon}) indeed defines the quasi--classical limit.

Since $\vec{E}^{a 2}$ is proportional to the action density $\rho_4$, and the typical time 
is $\Delta t \sim 1/k_{\perp}$, using (\ref{class}) we 
finally get that the classical description applies when
\beq 
k_{\perp}^2 < \alpha_s {N_g \over \pi R_A^2} \equiv Q_s^2.
\eeq

\subsection{The absence of mini--jet correlations}

When the occupation numbers of the field become large, the matrix elements of the creation and annihilation 
operators of the gluon field defined by
\beq
\hat{A}^{\mu} = \sum_{\vec{k}, \alpha} (\hat{c}_{\vec{k} \alpha} A^{\mu}_{\vec{k} \alpha} + 
\hat{c}^\dagger_{\vec{k} \alpha} A^{\mu *}_{\vec{k} \alpha})
\eeq
become very large,
\beq
N_{\vec{k} \alpha} = \langle  \hat{c}^\dagger_{\vec{k} \alpha} \hat{c}_{\vec{k} \alpha} \rangle \gg 1,
\eeq
so that one can neglect the unity on the r.h.s. of the commutation relation
\beq
\hat{c}_{\vec{k} \alpha}   \hat{c}^\dagger_{\vec{k} \alpha} - 
\hat{c}^\dagger_{\vec{k} \alpha} \hat{c}_{\vec{k} \alpha} = 1
\eeq 
and treat these operators as classical $c-$numbers.
 
This observation, often used in condensed matter physics, especially in the theoretical 
treatment of superfluidity, has important consequences for gluon production -- in particular, it implies that 
the correlations among the gluons in the saturation region can be neglected:
\beq
\langle A(k_1) A(k_2) ... A(k_n) \rangle \simeq \langle A(k_1) \rangle  \langle A(k_2) \rangle ... 
\langle A(k_n) \rangle. \label{abscor}
\eeq
Thus, in contrast to the perturbative picture, where the produced mini-jets have strong back-to-back 
correlations, the gluons resulting from the decay of the 
classical saturated field are uncorrelated at $k_{\perp} \lsim Q_s$.

Note that the amplitude with the factorization property (\ref{abscor}) is called 
point--like. However, the relation (\ref{abscor}) cannot be exact if we consider 
the correlations of final--state hadrons -- the gluon mini--jets cannot transform 
into hadrons independently. These correlations caused by color confinement however 
affect mainly hadrons with close three--momenta, as opposed to the perturbative 
correlations among mini--jets with the opposite three--momenta.

It will be interesting to explore the consequences of the factorization 
property of the classical gluon field (\ref{abscor}) for the HBT correlations of 
final--state hadrons. It is likely that the HBT radii in this case reflect  
the universal color correlations in the hadronization process.  

\vskip0.3cm

Another interesting property of classical fields follows from the relation 
\beq
\langle  (\hat{c}^\dagger_{\vec{k} \alpha} \hat{c}_{\vec{k} \alpha})^2 \rangle -
\langle  \hat{c}^\dagger_{\vec{k} \alpha} \hat{c}_{\vec{k} \alpha} \rangle^2 = 
 \langle  \hat{c}^\dagger_{\vec{k} \alpha} \hat{c}_{\vec{k} \alpha} \rangle, \label{fluctc}
\eeq
which determines the fluctuations in the number of produced gluons. 
We will discuss the implications of Eq. (\ref{fluctc}) for the multiplicity fluctuations 
in heavy ion collisions later.

\section{Classical QCD in action}

\subsection{Centrality dependence of hadron production}

In nuclear collisions, the saturation scale becomes a function of centrality; 
a generic feature of the quasi--classical 
approach -- the proportionality of the number of gluons to the inverse 
of the coupling constant (\ref{numb}) -- thus leads to definite predictions \cite{KN} 
on the centrality dependence of multiplicity. 

Let us first present the argument on a qualitative level.
At different centralities 
(determined by the impact parameter of the collision), the average density of partons 
(in the transverse plane)  
participating in the collision is very different. This density $\rho$ 
is proportional to the average length of nuclear material involved in the collision, 
which in turn approximately scales with the power of the number $N_{part}$ 
of participating nucleons, $\rho \sim N_{part}^{1/3}$.
The density of partons defines the value of the saturation scale, and so we expect
\be
Q_s^2 \sim N_{part}^{1/3}.
\ee
The gluon multiplicity is then, as we discussed above, is
\be
{dN_g \over d \eta} \sim {S_A \ Q_s^2 \over \alpha_s(Q_s^2)}, \label{mults}
\ee
where $S_A$ is the nuclear overlap area, 
determined by atomic number and the centrality of collision. Since $S_A\ Q_s^2 \sim N_{part}$ 
by definitions of the transverse density and area, from (\ref{mults}) we get
\be
{dN_g \over d \eta} \sim N_{part}\ \ln N_{part}, \label{mults1}
\ee
which shows that the gluon multiplicity shows a logarithmic deviation from the scaling 
in the number of participants.

To quantify the argument, we need to explicitly evaluate the average density of 
partons at a given centrality. This can be done by using Glauber theory, which 
allows to evaluate the differential cross section of the nucleus--nucleus interactions. 
The shape of the multiplicity distribution at a given (pseudo)rapidity $\eta$ can 
then be readily obtained (see, e.g., \cite{KLNS}):
\be
\frac {d \sigma} {d n} = \int d^2b \ {\cal P}(n;b)\ (1 - P_0(b)),  
\ee
where $P_0(b)$ is the probability of no interaction among the nuclei at a given 
impact parameter $b$: 
\be
P_0(b) = (1 - \sigma_{NN} T_{AB}(b))^{AB}; 
\ee
$\sigma_{NN}$ is the inelastic nucleon--nucleon cross section, and $T_{AB}(b)$ is the 
nuclear overlap function for the collision of nuclei with atomic numbers A and B; 
we have used the three--parameter Woods--Saxon nuclear density distributions \cite{tables}.

The correlation function ${\cal P}(n;b)$ is given by
\be
{\cal P}(n;b) = \frac{1}{\sqrt{2\pi a \bar{n}(b)}}\ \exp\left( - \frac{(n - \bar{n}(b))^2}{2 a 
\bar{n}(b)}\right),
\ee
here $\bar{n}(b)$ is the mean multiplicity at a given impact parameter $b$; 
the formulae for the number of participants and the number of binary 
collisions can be found in \cite{KLNS}.  The parameter 
$a$ describes the strength of fluctuations; for the classical gluon field, as follows from (\ref{fluctc}), 
$a = 1$. However, the strength of fluctuations can be changed by the subsequent evolution of the system 
and by hadronization process. Moreover, in a real experiment, the strength of fluctuations strongly 
depends on the acceptance. In describing the PHOBOS distribution \cite{PHOBOS}, we have found that the value $a = 0.6$ 
fits the data well. 
 
In Fig.4, we compare the 
resulting distributions for two different assumptions about the scaling of multiplicity with the number 
of participants to the PHOBOS 
experimental distribution, measured in the interval $3 < |\eta| < 4.5$.  
One can see that almost independently of theoretical assumptions about the dynamics of multiparticle production, 
the data are described quite well. At first this may seem surprising; the reason for this result is that at high energies, 
heavy nuclei are almost completely ``black''; unitarity then implies that  the shape of the cross section 
is determined almost entirely by the nuclear geometry.   
We can thus use experimental differential cross sections as 
a reliable handle on centrality. This gives us a possibility to compute the dependence of the 
saturation scale on centrality of the collision, and thus to predict the centrality dependence of 
particle multiplicities, shown in Fig. 5.  (see \cite{KN} for details).

\begin{figure}[htbp] 
\begin{center}
\epsfig{file=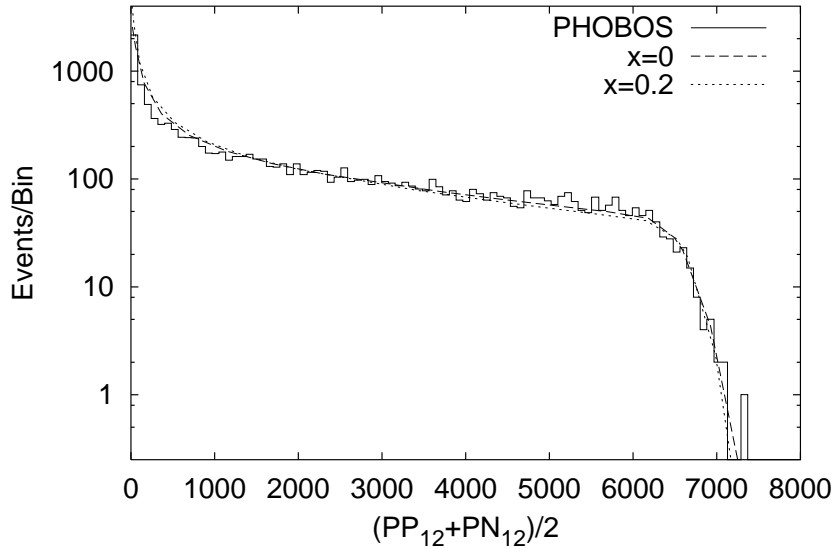,width=11cm}
\end{center}
\caption{Charged multiplicity distribution at $\sqrt{s}=130$ A GeV; solid line (histogram) -- 
PHOBOS result; dashed line -- distribution corresponding to 
participant scaling ($x=0$); 
dotted line -- distribution corresponding to the $37 \%$ admixture of ``hard'' component 
in the multiplicity; see 
text for details.}
\label{fig1}
\end{figure}

\begin{figure}[htbp]
\epsfig{file=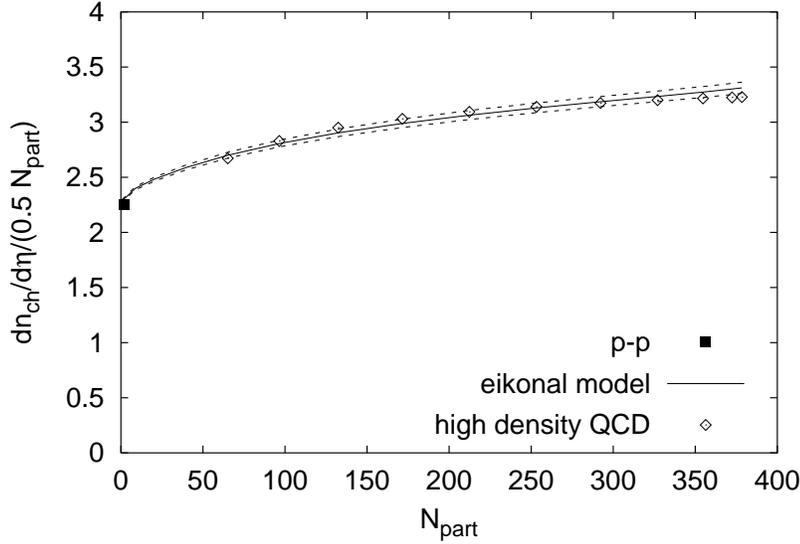,width=11cm}
%}
%\end{sideways}
\vskip0.1cm
\caption{Centrality dependence of the charged multiplicity per participant pair near $\eta = 0$ at 
$\sqrt{s} = 130$ A GeV; the curves represent the prediction based on the conventional eikonal 
approach, while the diamonds correspond to the high density QCD prediction (see text). The square 
indicates the $pp$ multiplicity.}
\label{centr}
\end{figure} 
    
%\begin{figure}[htbp]
%\begin{turn}{-90} 
%\epsfig{file=satscale.eps,width=11cm}
%\end{turn}
%\vskip0.1cm
%\caption{Saturation scale $Q_s^2$ dependence on the number of participants near $\eta = 0$ at 
%$\sqrt{s} = 130$ A GeV.}
%\label{centr}
%\end{figure}    

\subsection{Energy dependence}

\begin{figure}[h]
\begin{minipage}[t]{140mm}
\begin{turn}{-90}
\epsfig{file=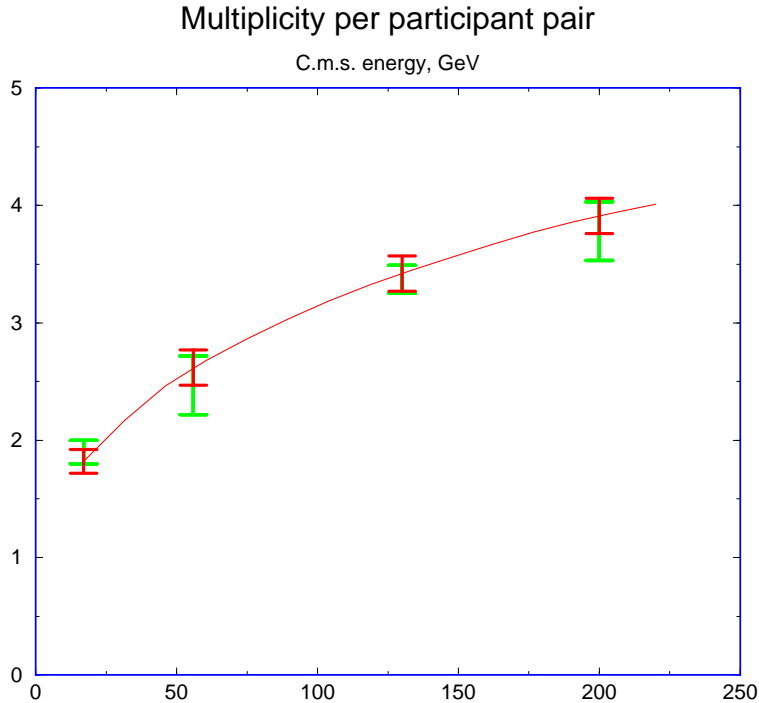,width=22pc}
\end{turn}
\caption{Energy dependence of charged multiplicity per participant pair at RHIC energies; solid line is the result 
(\ref{endep}).}
\label{fig:mult}
\end{minipage}
\end{figure}

Let us now turn to the discussion of energy dependence of hadron production. In semi--classical scenario, 
it is determined by the variation of saturation scale 
$Q_s$ with Bjorken $x = Q_s / \sqrt{s}$. This variation, in turn, is determined by the 
$x-$ dependence of the gluon structure function. 
In the saturation approach, the gluon distribution is related to the saturation scale 
by Eq.(\ref{qsat}). 
A good description of HERA data is obtained with saturation scale $Q_s^2 = 1 \div 2\ \rm{GeV}^2$
with $W$ - dependence ($W \equiv \sqrt{s}$ is the center-of-mass energy available 
in the photon--nucleon system)  \cite{GW} 
\beq
Q^2_s \,\,\propto\, W^{\lambda},
\eeq
where $\lambda \simeq 0.25 \div 0.3$. In spite of significant uncertainties in the determination 
of the gluon structure functions, perhaps even more important is the observation \cite{GW} that the 
HERA data exhibit scaling when plotted as a function of variable 
\beq
\tau \,=\, {Q^2 \over Q_0^2} \ \left({x \over x_0}\right)^{\lambda}, 
\eeq
where the value of $\lambda$ is again within the limits $\lambda \simeq 0.25 \div 0.3$. 
In high density QCD, this scaling is a consequence 
of the existence of dimensionful scale \cite{GLR,MV}) 
\beq
Q_s^2(x) = Q_0^2 \ (x_0 / x)^{\lambda}. 
\eeq
Using the value of $Q_s^2 \simeq 2.05\ {\rm GeV}^2$ extracted \cite{KN} at $\sqrt{s} = 130$ GeV and $\lambda = 0.25$ 
\cite{GW} used in \cite{KL}, equation (\ref{finres}) leads to the following approximate formula  
for the energy dependence of charged multiplicity in central $Au-Au$ collisions:
$$
\left<{2 \over N_{part}}\ {d N_{ch} \over d \eta}\right>_{\eta < 1} \approx 0.87\ 
\left({\sqrt{s}\ ({\rm GeV}) \over 130}\right)^{0.25}\ \times \nonumber
$$
\be
\times \left[3.93 + 0.25\ \ln\left({\sqrt{s}\ ({\rm GeV}) \over 130}\right)
\right]. \label{endep}
\ee
At $\sqrt{s} = 130\ {\rm GeV}$, we estimate from Eq.(\ref{endep}) 
$2/N_{part}\ dN_{ch}/d\eta \mid_{\eta<1} = 3.42 \pm 0.15$, 
to be compared to the average experimental value of $3.37 \pm 0.12$ 
\cite{PHOBOS,Phenix,Star,Brahms}. 
At $\sqrt{s} = 200\ {\rm GeV}$, one gets $3.91 \pm 0.15$, 
to be compared to the PHOBOS value \cite{PHOBOS} of $3.78 \pm 0.25$. 
Finally, at $\sqrt{s} = 56\ {\rm GeV}$, we find $2.62 \pm 0.15$, 
to be compared to \cite{PHOBOS} $2.47 \pm 0.25$. 
It is interesting to note that formula (\ref{endep}), when extrapolated to very high energies, 
predicts for the LHC energy a value 
substantially smaller than found in other approaches:
\be
\left<{2 \over N_{part}}\ {d N_{ch} \over d \eta}\right>_{\eta < 1} = 10.8 \pm 0.5; \ \ \  \sqrt{s} = 5500\ {\rm GeV},
\ee
corresponding only to a factor of $2.8$ increase in multiplicity between the RHIC energy of  $\sqrt{s} = 200\ {\rm GeV}$ 
and the LHC energy of $\sqrt{s} = 5500\ {\rm GeV}$  
(numerical calculations show that when normalized to the number of participants, 
the multiplicity in central $Au-Au$ and $Pb-Pb$ systems is almost identical).
The energy dependence of charged hadron multiplicity per participant pair is shown in Fig.\ref{fig:mult}.

One can also try to extract the value of the exponent $\lambda$ from the energy dependence of hadron 
multiplicity measured by PHOBOS at $\sqrt{s} = 130\ \rm{GeV}$ and at at $\sqrt{s} = 56\ \rm{GeV}$; 
this procedure yields $\lambda \simeq 0.37$, which is larger than 
the value inferred from the HERA data (and is very close to the value $\lambda \simeq 0.38$, 
resulting from the final--state saturation calculations \cite{EKRT}).

\subsection{Radiating the classical glue}

\begin{figure}[h]
\begin{minipage}{9.5cm}   
\begin{center}
\epsfysize=9.4cm
\leavevmode
\hbox{ \epsffile{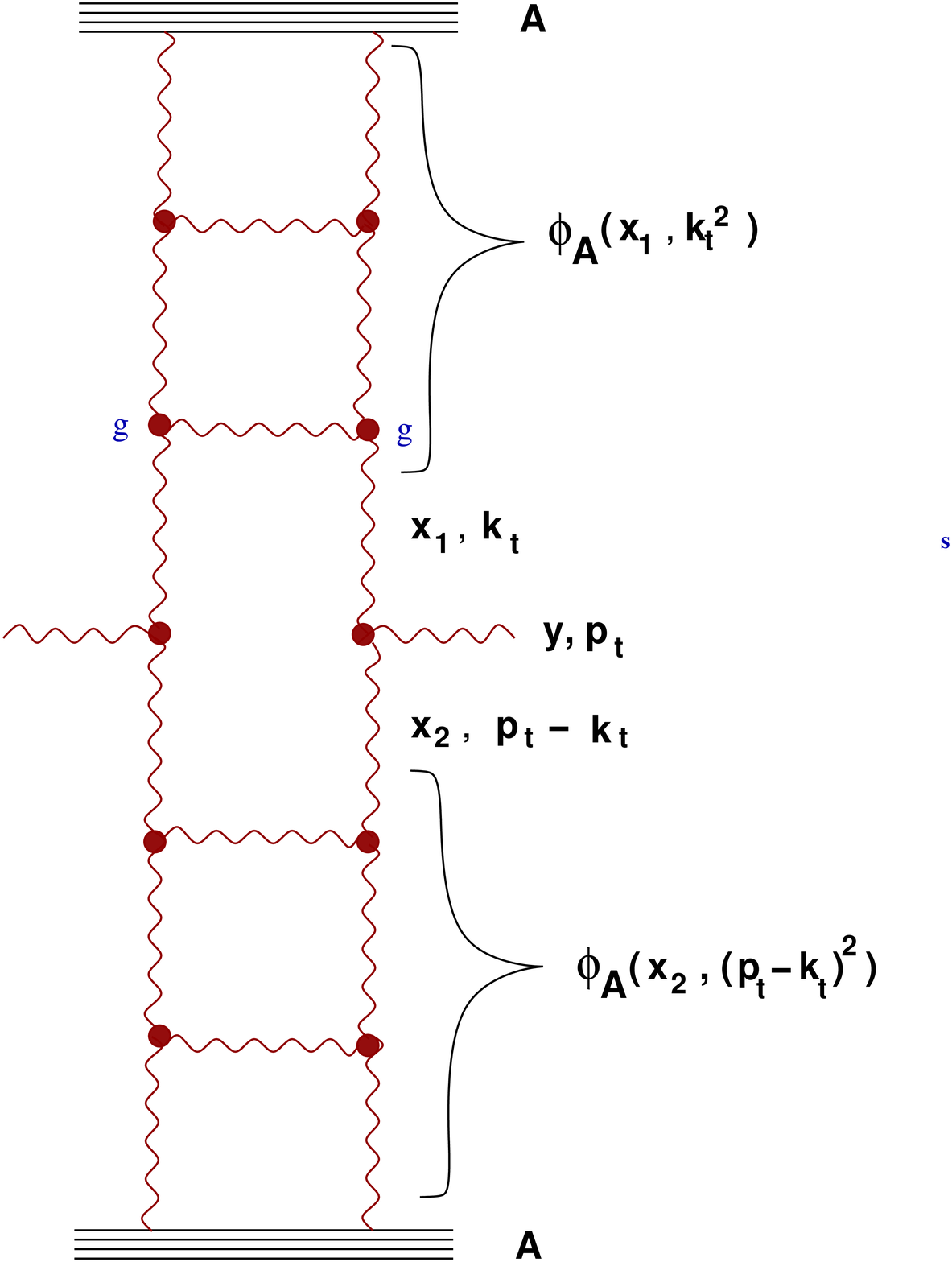}}
\end{center}
\end{minipage}
\begin{minipage}{5.5cm}
\caption{ The Mueller diagram for the classical gluon radiation.}
\label{phi}
\end{minipage}
\end{figure}

Let us now proceed to the quantitative calculation of the (pseudo-) 
rapidity and centrality dependences 
\cite{KL1}. 
We need to evaluate the leading tree diagram describing 
emission of gluons on the classical level, see Fig. \ref{phi}\footnote{Note that this ``mono--jet'' 
production diagram makes obvious the absence of azimuthal correlations in the saturation regime 
discussed above, see eq (\ref{abscor}).}.

%%%%%%%%%%%%%%%%%%%%%%%%%%%%%%%%%%%%%%%%%%%%%%%%%%%%%%%
Let us introduce the unintegrated gluon distribution $\varphi_A (x, k_t^2)$ which 
describes the probability to find a gluon with a given $x$ and transverse 
momentum $k_t$ inside the nucleus $A$. As follows from this definition, 
the unintegrated distribution is related to the gluon structure function by
\beq
xG_A(x, p_t^2) = \int^{p_t^2} d k_t^2 \ \varphi_A(x, k_t^2);
\eeq
when $p_t^2 > Q_s^2$, the unintegrated distribution corresponding to the bremsstrahlung 
radiation spectrum is 
\beq
\varphi_A(x, k_t^2) \sim {\alpha_s \over \pi} \ {1 \over k_t^2}.
\eeq 
In the saturation region, the gluon structure function is given by 
(\ref{glustr}); the corresponding unintegrated gluon distribution has only logarithmic dependence on the 
transverse momentum: 
\beq
\varphi_A(x, k_t^2) \sim {S_A \over \alpha_s}; \ k_t^2 \leq Q_s^2, \label{unint}
\eeq
where $S_A$ is the nuclear overlap area, determined by the atomic numbers of the 
colliding nuclei and by centrality of the collision.

 The differential cross section 
of gluon production in a $AA$ collision can now be written down as \cite{GLR,GM}
\beq
E {d \sigma \over d^3 p} = {4 \pi N_c \over N_c^2 - 1}\ {1 \over p_t^2}\ \int d k_t^2 \ 
\alpha_s \ \varphi_A(x_1, k_t^2)\ \varphi_A(x_2, (p-k)_t^2), \label{gencross}    
\eeq
where $x_{1,2} = (p_t/\sqrt{s}) \exp(\pm \eta)$, with $\eta$ the (pseudo)rapidity of the 
produced gluon; the running coupling $\alpha_s$ has to be evaluated at the 
scale $Q^2 = max\{k_t^2, (p-k)_t^2\}$. 
The rapidity density is then evaluated from (\ref{gencross}) according to 
\beq
{dN \over d y} = {1 \over \sigma_{AA}}\ \int d^2 p_t \left(E {d \sigma \over d^3 p}\right), 
\label{rapden}
\eeq
where $\sigma_{AA}$ is the inelastic cross section of nucleus--nucleus interaction.

Since the rapidity $y$ and Bjorken variable are related by $\ln 1/x = y$, 
the $x-$ dependence of the gluon structure function translates into the following 
dependence of the saturation scale $Q_s^2$ on rapidity:
\beq
Q_s^2(s; \pm y) = Q_s^2(s; y = 0)\ \exp(\pm \lambda y). \label{qsy}
\eeq

As it follows from (\ref{qsy}), the increase of rapidity at a fixed $W \equiv \sqrt{s}$ 
moves the wave function of one of the colliding 
nuclei deeper into the saturation region, while leading to a 
smaller gluon density in the other, which as a result can be 
pushed out of the saturation domain. Therefore, depending on the value of rapidity, 
the integration over the transverse momentum in Eqs. (\ref{gencross}),(\ref{rapden}) can be split in 
two regions: i) the region $\Lambda_{QCD} < k_t < Q_{s,min}$ in which the wave 
functions are both 
in the saturation domain; and ii) the region  $\Lambda << Q_{s,min} < k_t < Q_{s,max}$ in which 
the wave function of 
one of the nuclei is in the saturation region and the other one is not. 
Of course, there is 
also the region of $k_t > Q_{s,max}$, which is governed by the usual perturbative dynamics, 
but our assumption here is that the r{\^o}le of these genuine hard processes in the bulk 
of gluon production is relatively small; in the saturation scenario, 
these processes represent quantum fluctuations above the classical background. It is worth 
commenting that in the conventional mini--jet picture, this classical background is absent, 
and the multi--particle production is dominated by perturbative processes. 
This is the main physical difference between the two approaches; for the production 
of particles with $p_t >> Q_s$ they lead to identical results.    

To perform the calculation according to (\ref{rapden}),(\ref{gencross}) away from $y=0$ we need also 
to specify the behavior of the gluon structure function at large Bjorken $x$ (and out of 
the saturation region).  
At $x \to 1$, this behavior is governed by the QCD counting rules, $xG(x) \sim (1-x)^4$, so 
we adopt the following conventional form: $xG(x) \sim x^{-\lambda}\ (1-x)^4$.
  
We now have everything at hand to perform the integration over transverse momentum 
in (\ref{rapden}), (\ref{gencross}); the result is the 
following \cite{KL1}:
$$ 
{dN \over d y} = const\ S_A\ Q_{s,min}^2 \ \ln\left({Q_{s,min}^2 \over \Lambda_{QCD}^2}\right)
\ \times
$$
\beq
\times \ \left[1 + {1 \over 2}\ \ln\left({Q_{s,max}^2 \over Q_{s,min}^2}\right)\ 
\left(1 - {Q_{s,max} \over \sqrt{s}} e^{|y|}\right)^4\right],
 \label{resy}    
\eeq 
where the constant is energy--independent, $S_A$ is the nuclear overlap area, 
$Q_s^2 \equiv Q_s^2(s; y = 0)$, and $Q_{s,min(max)}$ 
are defined as the smaller (larger) values of (\ref{qsy}); at $y=0$, 
$Q_{s,min}^2 = Q_{s,max}^2 = Q_s^2(s) = Q_s^2(s_0)\ \times$ $\times (s /s_0)^{\lambda / 2}$. 
The first term in the brackets in (\ref{resy}) originates from the region in which both 
nuclear wave functions are in the saturation regime; this corresponds to 
the familiar $\sim (1/\alpha_s)\ Q_s^2 R_A^2$ term in the gluon multiplicity. 
The second term comes from the region in which only one of the wave functions is in 
the saturation region. The coefficient $1/2$ in front of the second term in 
square brackets comes from $k_t$ ordering of gluon momenta in evaluation of 
the integral of Eq.(\ref{gencross}).

The formula (\ref{resy}) has been derived using the form (\ref{unint}) 
for the unintegrated gluon distributions. We have checked numerically that the use of 
more sophisticated functional form of  
$\varphi_A$ taken from the saturation model of Golec-Biernat and W{\"u}sthoff \cite{GW} 
in Eq.(\ref{gencross}) affects the results only at the level of about $3\%$.

\vskip0.3cm

Since $S_A Q_s^2 \sim N_{part}$ (recall that $Q_s^2 \gg  \Lambda_{QCD}^2$ is defined 
as the density of partons 
in the transverse plane, which is proportional to the density of participants), we can 
re--write (\ref{resy}) in the following final form \cite{KL1} 
$$
{dN \over d y} = c\ N_{part}\ \left({s \over s_0}\right)^{\lambda \over 2}\ e^{- \lambda |y|}\ 
\left[\ln\left({Q_s^2 \over \Lambda_{QCD}^2}\right) - \lambda |y|\right]\ \times
$$
\beq
\times \left[ 1 +  \lambda |y| \left( 1 - {Q_s \over \sqrt{s}}\ e^{(1 + \lambda/2) |y|} \right)^4 
\right],  
\label{finres}
\eeq
with $Q_s^2(s) = Q_s^2(s_0)\ (s /s_0)^{\lambda / 2}$.
This formula is the central result of our paper; it expresses the predictions of 
high density QCD for the energy, centrality, rapidity, and atomic number dependences 
of hadron multiplicities in nuclear collisions in terms of a single scaling function. 
Once the energy--independent constant $c \sim 1$ and $Q_s^2(s_0)$ are determined 
at some energy $s_0$, Eq. (\ref{finres}) contains no free parameters. 
At $y = 0$ the expression (\ref{resy}) coincides exactly with the one 
derived in \cite{KN}, and extends it to describe the rapidity and energy dependences.  

\subsection{Converting gluons into hadrons}

The distribution (\ref{finres}) refers to the radiated gluons, while what is measured in experiment 
is, of course, the distribution of final hadrons. We thus have to make an assumption about the 
transformation of gluons into hadrons. The gluon mini--jets are produced with a certain virtuality, 
which changes as the system evolves; the distribution in rapidity is thus not preserved. 
However, in the analysis of jet structure it has been found that the {\it angle} 
of the produced gluon is remembered by the resulting 
hadrons; this property of ``local parton--hadron duality'' (see \cite{Yuri} and references therein) 
is natural if one assumes 
that the hadronization is a soft process which cannot change the direction of the emitted radiation.  
Instead of the distribution in the angle $\theta$, it is more convenient to use the distribution in 
pseudo--rapidity $\eta = - \ln \tan (\theta /2)$.  
Therefore, before we can compare (\ref{resy}) to the data, we have to convert the rapidity distribution 
(\ref{finres}) into the gluon distribution in pseudo--rapidity. We will then assume that the gluon and hadron 
distributions are dual to each other in the pseudo--rapidity space. 

To take account of the 
difference between rapidity $y$ and the measured pseudo-rapidity $\eta$, we have to multiply (\ref{resy}) 
by the Jacobian of the $y \leftrightarrow \eta$ transformation;
a simple calculation yields
\beq    
h(\eta; p_t; m) = \frac{\cosh \eta}{\sqrt{\frac{  m^2  \,+\,p_t^2}{p_t^2}\,\,+\,\,\sinh^2 
\eta}}, \label{Jac}
\eeq
where $m$ is the typical mass of the produced particle, and $p_t$ is its typical transverse 
momentum.
Of course, to plot the distribution (\ref{finres}) as a function of pseudo-rapidity, one also 
has to express rapidity $y$ in terms of pseudo-rapidity $\eta$; 
this relation is given by 
\beq
y (\eta; p_t; m) = {1 \over 2}\  \ln\left[{{\sqrt{\frac{  m^2  \,+\,p_t^2}{p_t^2}\,\,+\,\,\sinh^2 \eta}} + \sinh \eta} \over 
{{\sqrt{\frac{  m^2  \,+\,p_t^2}{p_t^2}\,\,+\,\,\sinh^2 \eta}} - \sinh \eta}\right]; \label{yeta}
\eeq
obviously, $h(\eta; p_t; m) = {\partial y (\eta; p_t; m) / \partial \eta}$. 

We now have to make an assumption about the typical invariant mass $m$ of the gluon mini--jet. 
%%%%%%%%%%%%%%%%%%%%%%%%%%%%%%%%%%%%%%%%%%%%%%%%%%%%%%%%%%%%%%%%%%%%%%%%%
Let us estimate it by assuming that the slowest hadron in the mini-jet decay is 
the $\rho$-resonance, with energy $E_{\rho} = (m_{\rho}^2 + p_{\rho,t}^2 + p_{\rho,z}^2)^{1/2}$, 
where the $z$ axis is pointing along the mini-jet momentum.  
Let us also denote by $x_i$ the fractions of the gluon energy $q_0$ 
carried by other, fast, $i$ particles in the mini-jet decay. Since the sum of transverse (with respect 
to the mini-jet axis) momenta 
of mini-jet decay products is equal to zero, the 
mini-jet invariant mass $m$ is given by 
$$
m^2_{jet}\,\equiv\, m^2 = ( \sum_i x_i q_0 + E_{\rho})^2 - (\sum_i x_i
q_z + p_{\rho,z})^2 \,\simeq\,\,
$$
\begin{equation}
\label{decay}
 \,\simeq\,\,
2 \sum_i x_i  q_z \cdot ( m_{\rho,t} - p_{\rho,z}) \equiv 2 Q_s \cdot m_{eff},
\end{equation}
where $m_{\rho,t}=(m_{\rho}^2 + p_{\rho,t}^2)^{1/2}$.
In \eq{decay} we used that $\sum_i x_i =1$ and $q_0 \approx q_z = Q_s$.
Taking $p_{\rho,z} \approx p_{\rho,t}  \approx\,300$ MeV and $\rho$ mass, we obtain $m_{eff}
\approx 0.5\,$ GeV.

We thus use the mass $m^2 \simeq 2 Q_s m_{eff} \simeq Q_s \cdot 1$ GeV in Eqs.(\ref{Jac},\ref{yeta}).
 Since the typical transverse momentum of the produced 
gluon mini--jet is $Q_s$, we take $p_t = Q_s$ in (\ref{Jac}).
The effect of the 
transformation from rapidity to pseudo--rapidity is the decrease of multiplicity at 
small $\eta$ by 
about $25-30 \%$, leading to the appearance of the $\approx 10 \%$ dip in the pseudo--rapidity 
distribution in the 
vicinity of $\eta = 0$. 
 We have checked that the change in the value of the mini--jet mass by two times affects the 
Jacobian at central pseudo--rapidity to about $\simeq 10\%$, leading to $\sim 3\%$ effect on the 
final result.

\begin{figure}[htbp] 
\begin{center}
\epsfig{file=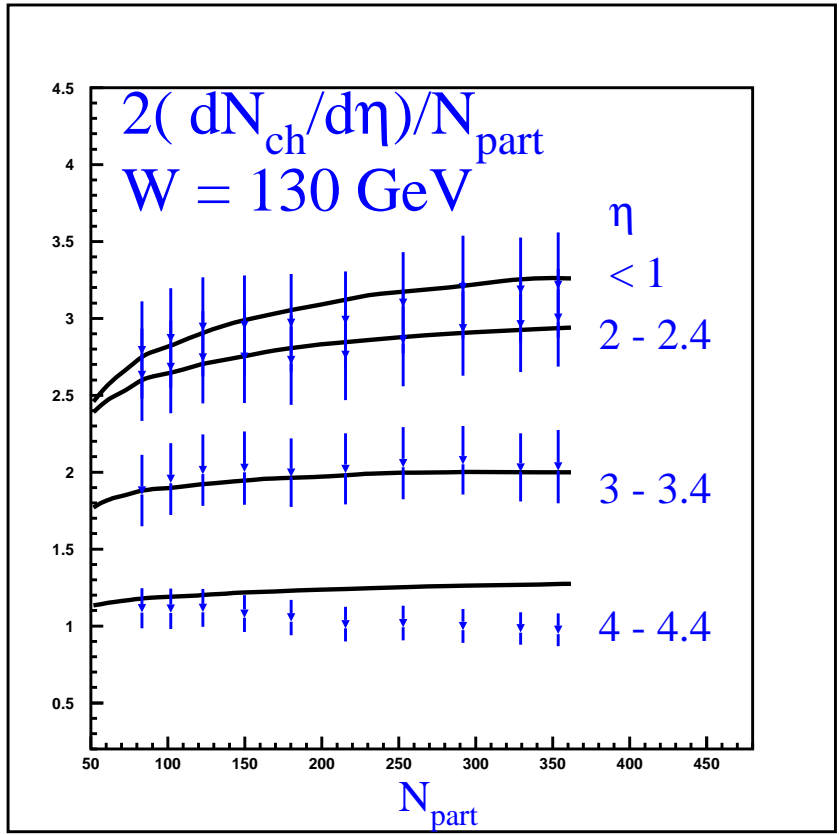,width=11cm}
\end{center}
\caption{Centrality dependence of charged hadron production per participant at different 
pseudo-rapidity $\eta$ intervals in $Au-Au$ collisions 
at $\sqrt{s} = 130$ GeV; from (Kharzeev and Levin, 2001), the data are from (PHOBOS Coll., 2000).}
\label{fig1x}
\end{figure}

\begin{figure}[htbp] 
\begin{center}
\epsfig{file=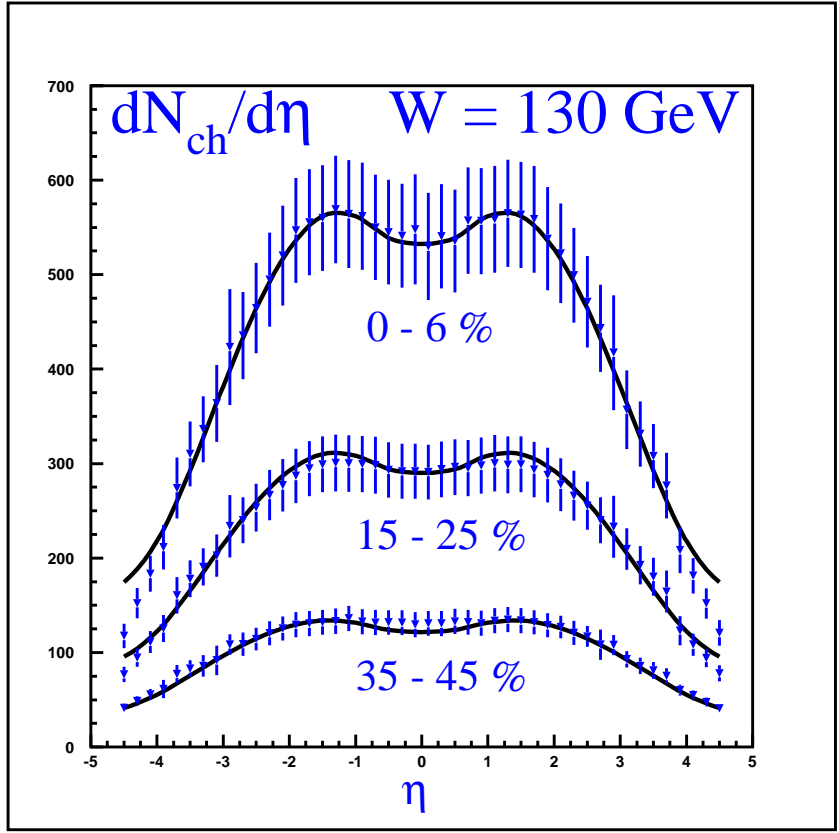,width=11cm}
\end{center}
\caption{ Pseudo--rapidity dependence of charged hadron production at different cuts on centrality 
in $Au-Au$ collisions 
at $\sqrt{s} = 130$ GeV; from (Kharzeev and Levin, 2001), the data are from (PHOBOS Coll., 2000).}
\label{fig2}
\end{figure}

The results for the $Au-Au$ collisions at $\sqrt{s} = 130$ GeV are presented in Figs \ref{fig1x} 
and \ref{fig2}. In the calculation, we use the results on the dependence 
of saturation scale on the mean number of participants at $\sqrt{s} = 130$ GeV 
from \cite{KN}, see Table 2 of that 
paper. The mean number of participants in a given centrality cut is taken from the PHOBOS paper 
\cite{PHOBOS}. One can see that both the centrality dependence 
 and the rapidity dependence of the $\sqrt{s} = 130$ GeV PHOBOS data are well reproduced 
below $\eta \simeq \pm 4$. The rapidity dependence has been evaluated 
with $\lambda = 0.25$, which is within the 
range $\lambda = 0.25 \div 0.3$ inferred from the HERA data \cite{GW}. The discrepancy 
above $\eta \simeq \pm 4$ is not surprising since 
our approach does not properly take into account multi--parton correlations 
which are important in the fragmentation region.

Our predictions for $Au-Au$ collisions at $\sqrt{s} = 200$ GeV are presented in \cite{KL1}.
The only parameter which governs the energy dependence is the exponent $\lambda$, which we 
assume to be $\lambda \simeq 0.25$ as inferred from the HERA data. The absolute 
prediction for the multiplicity, as explained above, bears some uncertainty, but there is a definite 
feature of our scenario which is distinct from other approaches. 
It is the dependence of multiplicity on centrality, which around $\eta =0$ is determined 
solely by the running of the QCD strong coupling \cite{KN}. As a result, the centrality dependence 
at $\sqrt{s} = 200$ GeV is somewhat less steep than at $\sqrt{s} = 130$. While the 
difference in the shape at these two energies is quite small, in the perturbative 
mini-jet picture this slope 
should increase, reflecting the growth of the mini-jet cross section with 
energy \cite{WG}. 

\subsection{Further tests}

Checking the predictions of the semi--classical approach for the centrality and pseudo--rapidity 
dependence at $\sqrt{s} = 200$ GeV is clearly very important. What other tests 
of this picture can one devise? The main feature of the classical emission is that it is 
coherent up to the transverse momenta of about $\sqrt{2}\ Q_s$ (about $\simeq 2$ GeV/c for 
central $Au-Au$ collisions). This means 
that if we look at the centrality dependence of particle multiplicities above a certain value of the  
transverse momentum, say, above $1$ GeV/c, it should be very similar to the dependence 
without the transverse momentum cut-off. On the other hand, in the two--component ``soft plus hard'' model 
the cut on the transverse momentum would strongly enhance the contribution of hard 
mini--jet production processes, since soft production mechanisms presumably do not 
contribute to particle production at high transverse momenta.  
Of course, at sufficiently large value of the cutoff all of the observed particles 
will originate from genuine hard processes, and the centrality dependence will 
become steeper, reflecting the scaling with the number of collisions. It will be very 
interesting to explore the transition to this hard scattering regime experimentally. 

Another test, already discussed above (see eq.(\ref{abscor})) is the study of azimuthal 
correlations between the produced high $p_t$ particles. In the saturation scenario 
these correlations should be very small below $p_t \simeq 2$ GeV/c in central 
collisions. At higher transverse momenta, and/or for more peripheral collisions 
(where the saturation scale is smaller) these correlations should be much 
stronger.

\section{Does the vacuum melt?}

The approach described above allows us to estimate the initial energy density of partons 
achieved at RHIC. Indeed, in this approach the formation time of 
partons is $\tau_0 \simeq 1/Q_s$, and the transverse momenta of partons are about $k_t \simeq Q_s$. 
We thus can use the Bjorken formula and the set of parameters deduced above to estimate \cite{KN}
\be
\epsilon \simeq {<k_t> \over \tau_0} \ {d^2 N \over d^2b d \eta} \simeq Q_s^2 \ {d^2 N \over d^2b d \eta} \simeq 
18\ {\rm {GeV/fm^3}}
\ee
for central $Au-Au$ collisions at $\sqrt{s} = 130$ GeV.
This value is well above the energy density needed to induce the  
QCD phase transition according to the lattice calculations.  However, the picture of gluon production 
considered above seems to imply that the gluons simply flow from the initial state of the incident nuclei 
to the final state, where they fragment into hadrons, with 
nothing spectacular happening on the way. In fact, one may even wonder if the presence of these gluons 
modifies at all the structure of the physical QCD vacuum. 

To answer this question theoretically, we have to possess some knowledge about the non--perturbative 
vacuum properties. 
While in general the problem of vacuum structure still has not been solved (and this is one of the main 
reasons for the heavy ion research!), we do know one class of vacuum solutions -- the instantons.  
It is thus interesting to investigate what happens to the QCD vacuum in the presence of strong 
external classical fields using the example of instantons \cite{KKL}.

\begin{figure}
\begin{center}
\epsfxsize=12cm
\leavevmode
\hbox{ \epsffile{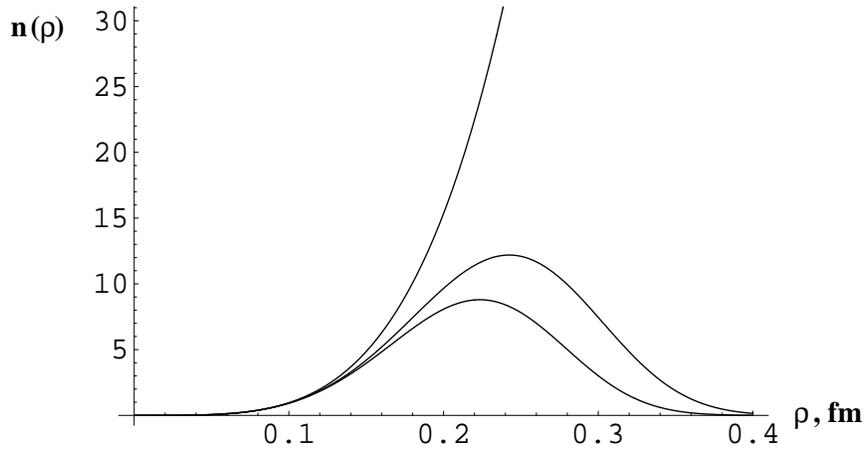}}
\end{center}
\caption{Distributions of instanton sizes in vacuum for QCD with 
three light flavors (upper curve) versus the distribution of instanton
sizes in the saturation environment produced by a collision of two
identical nuclei for $c=1$ (middle curve) and $c = 2 \ln 2$ (lower
curve) with $Q_s^2 \, = \, 2 \, \mbox{GeV}^2$; from (Kharzeev, Kovchegov and Levin, 2002).}
\label{nsat}
\end{figure}

The problem of small instantons in a slowly varying background field
was first addressed in \cite{cdg1,svz} by introducing
the effective instanton lagrangian $L^{I(\overline{I})}_{eff}
(x)$ 
\be\label{effl}
L_{eff}^I (x_0) \, = \, \int d \rho \, n_0 (\rho) \, dR \, \exp \left(
- \frac{2 \pi^2}{g} \, \rho^2 \, \overline{\eta}^M_{a\mu\nu} \, R^{a
a'} \, G^{a'}_{\mu\nu} (x_0) \right)
\ee
in which $n_0 (\rho)$ is the instanton size distribution function in the 
vacuum, $\overline{\eta}^M_{a\mu\nu}$ is the 't Hooft symbol in Minkowski space, 
and $R^{aa'}$ is the matrix of rotations in color space, with $dR$ denoting 
the averaging over the instanton color orientations. 

 The complete field of a single instanton solution could be
reconstructed by perturbatively resumming the powers of the effective
instanton lagrangian which corresponds to perturbation theory in
powers of the instanton size parameter $\rho^2$. In our case here the
background field arises due to the strong source current
$J_\mu^a$. The current can be due to a single nucleus, or resulting from the 
two colliding nuclei. Perturbative resummation of powers of
the source current term translates itself into resummation of the
powers of the classical field parameter $\as^2 A^{1/3}$
\cite{MV,yuri}. Thus the problem of instantons in the background 
classical gluon field is described by the effective action in
Minkowski space
\be\label{qcdi}
S_{eff} \, = \, \int d^4 x \left( - \frac{1}{4} G^a_{\mu\nu} (x)
G^a_{\mu\nu} (x) \, + \, L^I_{eff} (x) \, + \, L^{\overline{I}}_{eff}
(x) \, + \, J_\mu^a \, A_\mu^a (x) \right).
\ee

The problem thus is clearly formulated; by using an explicit form for the 
radiated classical gluon field, it was possible to demonstrate \cite{KKL} that
the distribution of
instantons gets modified from the original vacuum one $n_0(\rho)$ to
\be\label{naa}
n_{sat}^{AA} (\rho) \, = \, n_0 (\rho) \, \exp \left( - \frac{c \,
\rho^4 Q_s^4 } { 8 \, \as^2 \, N_c \, (Q_s \tau_0)^2 } \right),
\ee
where $\tau_0$ is the proper time.
The result \eq{naa} shows that large size instantons are 
suppressed by the strong classical fields generated in the nuclear
collision (see Fig. \ref{nsat})\footnote{Of course, at large proper times $\tau_0 \to \infty$ 
the vacuum ``cools off'', and the instanton distribution returns to the vacuum one.}.
The vacuum does melt!

%\section{Summary}

\vskip0.3cm

The results presented here were obtained together with Hirotsugu Fujii, Yuri Kovchegov, 
Eugene Levin, and Marzia Nardi. I am very grateful to them for the most enjoyable collaboration.
I also wish to thank Jean-Paul Blaizot, Yuri Dokshitzer, Larry McLerran, Al Mueller and Raju Venugopalan 
for numerous illuminating discussions on the subject of these lectures. 
This work was supported by the U.S. Department of Energy under Contract No. DE-AC02-98CH10886. 

\end{article}
\end{document}